\documentclass[prc,twocolumn,english,showpacs]{revtex4}
\usepackage[T1]{fontenc}
\usepackage[latin1]{inputenc}
\usepackage{babel}
\usepackage{graphics}
\usepackage{comment}

\makeatletter

\usepackage[T1]{fontenc}
\usepackage[latin1]{inputenc}
\usepackage{babel}
\usepackage{graphics}

\renewcommand {\vec}[1]{\mbox{\boldmath $#1$}}

\newcommand{\ra}{\rangle}
\newcommand{\la}{\langle}
\newcommand{\rp}{\right)}
\newcommand{\lp}{\left(}
\newcommand{\rc}{\right]}
\newcommand{\lc}{\left[}
\newcommand{\emax}{\varepsilon_\mathrm{max}}



\makeatother
\begin{document}

\title{Four-body continuum-discretized coupled-channels calculations using a transformed harmonic oscillator basis}

\author{M. Rodr\'{\i}guez-Gallardo$^{1,2,3}$, J.~M.~Arias$^2$,
J.~G\'omez-Camacho$^2$, R.~C.~Johnson$^3$, A.~M. Moro$^2$, I.~J. Thompson$^{3,4}$, and J.~A. Tostevin$^3$}

\affiliation {$^1$ Centro de Física Nuclear, Universidade de Lisboa, Av. Prof. Gama Pinto nº2, 1649-003 Lisboa, Portugal}
\affiliation {$^2$ Departamento de F\'{\i}sica At\'omica,
Molecular y Nuclear, \\ Universidad de
Sevilla, Apartado 1065, 41080 Sevilla, Spain} \affiliation{$^3$
Department of Physics, University of Surrey, Guildford GU2 7XH,
United Kingdom}
\affiliation{$^4$  Lawrence Livermore National Laboratory,
 PO Box 808, L-414,  Livermore, CA 94551, USA}

\date{\today}

\begin{abstract}
The scattering of a weakly bound three-body system by a target is
discussed. A transformed harmonic oscillator basis is used to provide
an appropriate discrete and finite basis for treating the continuum
part of the spectrum of the projectile. The continuum-discretized
coupled-channels framework is used for the scattering
calculations. The formalism is applied to different reactions,
$^6$He+$^{12}$C at 229.8 MeV, $^6$He+$^{64}$Zn at 10 and 13.6 MeV, and
$^6$He+$^{208}$Pb at 22 MeV, induced by the Borromean nucleus
$^6$He. Both the Coulomb and nuclear interactions with a target are
taken into account.
\end{abstract}
\maketitle

\section{Introduction}

The development of radioactive nuclear beam facilities has allowed the
study of nuclei far from the line of stability, bringing to the fore
new nuclear structure problems. A significant topic in recent years
has been the study of halo nuclei \cite{Rii94,Han95,Jen04}.  These are
weakly bound, spatially extended systems, typically comprising a core
and one or two valence nucleons.  Particularly interesting examples of
such systems are Borromean nuclei, i.e., three-body composite systems
with no binary bound states. These nuclei have attracted special
attention because their loosely bound nature reflects a delicate
interplay between two- and three-body forces, constituting a challenge
to existing theories, and a motivation for the development of new
ones. The detailed structure of the continuum spectrum of these
systems is still not fully understood, partially due to the
ambiguities associated with the underlying forces between the
constituents. Due to their low binding energy, halo nuclei are easily
broken up in the nuclear and Coulomb field of the target nucleus.
Therefore few-body reaction theories, developed to extract reliable
information from experimental data of reactions involving loosely
bound systems, have to include, as an essential ingredient, a
realistic description of coupling to the continuum part of the
spectrum.

From the theoretical point of view, the treatment of reactions
involving loosely bound systems must deal with the complication that
these continuum breakup states are not square-normalizable.  A
convenient method to circumvent this problem is to replace the states
in the continuum by a finite set of normalized states, thus providing
a discrete basis that, hopefully, can be truncated to a small number
of states and yet provide a reliable description of the
continuum. Several prescriptions to construct such a discrete basis
have been proposed.  For two-body composite systems, where the
continuum states are easily calculated, one can use a discretization
procedure in which the continuum spectrum is truncated at a maximum
excitation energy and divided into energy intervals. For each
interval, or bin, a normalizable state is constructed by superposition
of scattering states within that bin interval.  The method, normally
used in the continuum-discretized coupled-channels (CDCC) framework
\cite{Yah86,Aus87}, has been very useful in the description of elastic
and breakup observables in reactions involving weakly bound two-body
projectiles.

An alternative to the binning procedure is to represent the continuum
spectrum by the eigenstates of the internal Hamiltonian in a basis of
square integrable (or $L^2$) functions, such as Laguerre
\cite{Bra95,Kur82,Kur91}, Gaussian \cite{Mat03,Mac87a} or Sturmian
\cite{Joh74,And78,Wen80,Lai93} functions. In practice, the
diagonalization is performed in a finite (truncated) set of states and
the resulting eigenstates, also known as {\em pseudo-states} (PS), are
regarded as a finite and discrete representation of the spectrum of
the system.  The pseudo-states are then used within a coupled-channels
calculation in the same way as the continuum bins.

The PS method has the appealing feature of being readily applicable
also to describe the spectrum of three-body systems, in which case the
Hamiltonian is diagonalized in a complete set of square-integrable
functions for the three-body Hilbert space. Several applications of
this method can be found in the literature, for both structure
\cite{Hiy03} and reaction problems \cite{Mat04}.  In the latter case,
the method is an extension of the CDCC formalism to reactions with
three-body projectiles, using a pseudo-state model for the continuum.

One such PS method proposed recently is the Transformed Harmonic
Oscillator (THO) method \cite{Per01a,Per01b}.  Given the ground-state
wave function of the system, the THO method performs a Local Scale
Transformation (LST) \cite{Pet81} that converts the bound ground-state
wave function of the system into the ground-state wave function of a
Harmonic Oscillator (HO).  Once the LST is obtained, the HO basis can
be transformed, by the inverse LST, to a discrete basis in the
physical space. The THO basis functions are not eigenfunctions of the
Hamiltonian (except for the ground state) but the Hamiltonian can be
diagonalized in an appropriate truncated basis to produce approximate
eigenvalues and eigenfunctions. This method has been shown to be
useful for describing the two-body continuum in both structure
\cite{Per01a,Per01b,Rod04} and scattering \cite{Mor02,Mar02,Mor06}
problems.  In a recent work \cite{Rod05} the THO method was
generalized to describe continuum states of three-body systems, based
on expansion in Hyperspherical Harmonics (HH) \cite{Zhu93}.  In
particular the method was applied to the Borromean nucleus $^6$He, for
which several strength functions, including the dipole and quadrupole
Coulomb transition strengths, were calculated.  These observables are
found to converge quickly with respect to the number of THO basis
states included.  Furthermore, the calculated strength distributions
are in very good agreement with those obtained using three-body
scattering wave functions \cite{Tho00}.

Most of our knowledge of $^6$He comes from the analysis of reactions
where secondary beams collide with stable nuclei. These experiments
have been performed with both light \cite{Aks03,Ege02} and heavy
targets \cite{Aum98,Aum99,Agu01,Agu00,Kak03}, and at low and high
energies, providing a body of data which can be used to benchmark
reaction and structure models.  The theoretical understanding of
reactions involving a three-body projectile, such as $^6$He, is a
complicated task because it requires the solution of a four-body
scattering problem. At high energies, a variety of approximations have
been used such as semiclassical approximations
\cite{Glau59,Alk97,Alk98}, {\em frozen halo} or adiabatic
approximations \cite{Ron97b,Chr97}, Multiple Scattering expansions
\cite{Ker59,Wat57,Cres99}, four-body DWBA \cite{Chat01,Ers04}, among
others. However, at energies of a few MeV per nucleon, some of these
approximations are not justified.  Then the use of the CDCC method is
an alternative to solve these problems. For a four-body problem
(three-body projectile) this method has already been applied using a
PS basis based on Gaussian functions. The scattering of $^6$He by
$^{12}$C \cite{Mat04} and $^{209}$Bi \cite{Mat06} have been
studied. In both cases a good agreement was obtained with the
experimental data of Refs.~\cite{Lap02,Mil04} and \cite{Agu01},
respectively.

In this work, we study the scattering of a three-body projectile by a
target using the CDCC formalism. The novel feature of the present
approach is the use of the THO PS basis to represent the states of the
projectile. These states are then used to generate the
projectile-target coupling potentials that enter the system of coupled
equations.  Furthermore, we have developed a new procedure to
calculate these coupling potentials making use of an expansion of the
wave functions of the projectile internal states in a HH basis.

This paper is structured as follows. In Section II the three-body
discretization method is presented. In Section III the multipole
expansion of the interaction potential between the projectile and the
target is addressed. In Section IV we describe the three-body model
for the Borromean nucleus $^6$He. In Section V we apply the formalism
to the reactions $^6$He$+^{12}$C at $E_\mathrm{lab}$=229.8 MeV, 
$^6$He$+^{64}$Zn at $E_\mathrm{lab}$=13.6 and 10 MeV, 
and $^6$He$+^{208}$Pb at $E_\mathrm{lab}$=22 MeV.  Finally,
Section VI summarizes and draws conclusions.

\section{Three-body continuum discretization method}

The THO discretization method applied to a three-body system is
described in detail in Ref.~\cite{Rod05}. For completeness, in this
Section we outline the main features of the formalism.  In the
three-body case, it is convenient to work with the hyperspherical
coordinates $\{\rho,\alpha,\widehat{x},\widehat{y}\}$. They are
obtained from the Jacobi coordinates $\{\vec{x},\vec{y}\}$ that are
illustrated in Fig.~\ref{4c}. The variable \vec{x} is proportional to
the relative coordinate between two of the particles, with a scaling
factor depending on their masses ~\cite{Rod04} and \vec{y} is
proportional to the coordinate from the center of mass of these two
particles to the third particle, again with a scaling factor depending
on their masses. From these coordinates, the hyperradius ($\rho$) and
the hyperangle ($\alpha$) are defined as $\rho=\sqrt{x^2+y^2}$ and
$\tan{\alpha}=x/y$. Obviously there are three different Jacobi sets
but $\rho$ is the same for all of them.

For a three-body system the discretization method has two parts.
First, the wave functions of the system are expanded in Hyperspherical
Harmonics (HH) \cite{Zhu93}. We define states of good total angular
momentum as
\begin{eqnarray}
{\mathcal Y}_{\beta j\mu}(\Omega)&=& \sum_{\nu\iota}\la j_{ab}\nu
I\iota|j\mu\ra\chi_I^{\iota}\nonumber\\ &\times&\sum_{m_l\sigma}\la
lm_lS_x\sigma|j_{ab}\nu\ra\Upsilon_{Klm_l}^{l_xl_y}(\Omega)\chi_{S_x}^{\sigma},
\label{HH}
\end{eqnarray}
where $\Upsilon_{Klm}^{l_xl_y}(\Omega)$ are the hyperspherical
harmonics that depend on the angular variables
$\Omega\equiv\{\alpha,\widehat{x},\widehat{y}\}$,
$\chi_{S_x}^{\sigma}$ is the spin wave function of the two particles
related by the coordinate $\vec{x}$, and $\chi_I^{\iota}$ is the spin
function of the third particle. Each component of the wavefunction (or
channel) is defined by the set of quantum numbers
$\beta\equiv\{K,l_x,l_y,l,S_x,j_{ab}\}$. Here, $K$ is the
hypermomentum, $l_x$ and $l_y$ are the orbital angular momenta
associated with the Jacobi coordinates $\vec{x}$ and $\vec{y}$,
$\vec{l}=\vec{l}_x+\vec{l}_y$ is the total orbital angular momentum,
$S_x$ is the spin of the particles related by the coordinate
$\vec{x}$, and $\vec{j}_{ab}=\vec{l}+\vec{S}_x$. Finally,
$\vec{j}=\vec{j}_{ab}+\vec{I}$ is the total angular momentum, with $I$
the spin of the third particle, which we assume fixed.  The physical
states of the system can now be expressed as a linear combination of
the states given by Eq.~(\ref{HH}) as
\begin{equation}
\psi_{j\mu}(\rho,\Omega) =\sum_{\beta}R_{\beta
j}(\rho)\mathcal{Y}_{\beta j\mu}(\Omega),
\label{BHH}
\end{equation}
where $\{R_{\beta j}\}$ are the hyperradial wave functions.

Secondly, the THO method is used to obtain the functions $R_{\beta
j}(\rho)$.  Writing the ground-state wave function in the form of
Eq.~(\ref{BHH}), the equation that defines the LST for each channel
$\beta$ is
\begin{center}
\begin{equation}
\label{LST}
|N_{B\beta}|^2\int_0^{\rho}d\rho'\rho'^5|R_{B\beta}(\rho')|^2=\int_0^{s}ds's'^5|R_{0K}^{HO}(s')|^2,
\end{equation}
\end{center}
where $R_{B\beta}(\rho)$ is the bound ground-state hyperradial wave
function for the channel $\beta$, with $N_{B\beta}$ the normalization
factor, and $R_{0K}^{HO}(s)$ is the ground-state hyperradial wave
function of the HO for the hypermomentum $K$, that is already
normalized.  Finally, the THO basis is constructed for each channel by
applying the LST, $s_{\beta}(\rho)$, to the HO basis
\begin{eqnarray}
\label{rtho}
R^{THO}_{i\beta}(\rho)&=&\frac{{\mathcal{N}}_{iK}}{{\mathcal{N}}_{0K}}N_{B\beta}R_{B\beta}(\rho)L_i^{K+2}(s_{\beta}(\rho)^2),\\
\label{phitho}
\psi^{THO}_{i\beta
j\mu}(\rho,\Omega)&=&R^{THO}_{i\beta}(\rho)\mathcal{Y}_{\beta
j\mu}(\Omega)
\end{eqnarray}
where the $L_i^{\lambda}(t)$ are generalized Laguerre polynomials and
${\mathcal{N}}_{iK}$ is the normalization constant of the HO
basis. Here the index $i$ denotes the number of hyperradial
excitations. Note that as $i$ increases, the functions
$R^{THO}_{i\beta}(\rho)$ become more oscillatory and explore larger
distances.

For channels with quantum numbers that do not contribute to the
ground-state wave function, the (ground state) channel with the
closest quantum labels to the channel of interest is used to construct
the LST.  One important point concerns the label $K$ which governs the
$\rho^K$ behavior of the hyperradial wave function close to the
origin. To guarantee the correct behavior of the wavefunction, we
select a channel from the ground-state wave function with the same
$K$.  If this is not possible, a channel with $K-1$ is used and the
corresponding hyperradial wave function is then multiplied by $\rho$.

The required discrete eigenstates are now calculated by diagonalizing
the three-body Hamiltonian of the projectile in a finite THO basis up
to $n_b$ hyperradial excitations in each channel,
\begin{equation}
\phi_{nj\mu}^{THO}(\vec{x},\vec{y})=\sum_{\beta}\sum_{i=0}^{n_b}C_{n}^{i\beta
j}\psi_{i\beta j\mu}^{THO}(\rho,\Omega),
\label{foae-1}
\end{equation}
where $n$ labels the eigenstates for a given angular momentum $j$ and
$\varepsilon_{nj}$ will be the associated energy.  Replacing in this
expression the functions $\psi_{i\beta j\mu}^{THO}(\rho,\Omega)$ by
their explicit expansion in terms of the HH, Eq.~(\ref{phitho}), and
performing the sum in the index $i$ for $i=0,\ldots,n_b$, we can
express the PS basis states as
\begin{equation}
\phi^{THO}_{nj\mu}(\vec{x},\vec{y})=\sum_{\beta}R_{n \beta
j}^{THO}(\rho){\mathcal Y}_{\beta j\mu}(\Omega).
\label{foae}
\end{equation}

 Note that the choice of the HO parameter has no influence in the
 calculation of the LST since changes to this parameter are equivalent
 to making a linear transformation in the oscillator variable
 $s$. This gives the same result for the right part of
 Eq.~(\ref{LST}).

\section{Multipole expansion of the projectile-target potential}

\label{multexp}

 \begin{figure}
\resizebox{0.4\textwidth}{!}{%
  \includegraphics{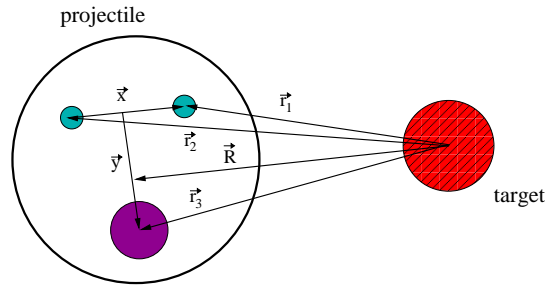}
}
\caption{(Color online) Relevant coordinates for the scattering of a three-body projectile by a structureless target.}
\label{4c}
\end{figure}

The eigenstates given in Eq.~(\ref{foae}) are a discrete
representation of the states of the three-body projectile. From them,
the four-body wavefunction of the projectile-target system,
schematically depicted in Fig.~\ref{4c}, is formed as
\begin{eqnarray}
\Psi_{JM}(\vec{R},\vec{x},\vec{y})&=&\sum_{nj\mu
LM_L}\phi_{nj\mu}^{THO}(\vec{x},\vec{y})\la LM_Lj\mu|JM\ra
i^L\nonumber\\ &\times&Y_{LM_L}(\widehat{R})\frac{1}{R}f_{Lnj}^J(R),
\end{eqnarray}
where $\vec{R}$ is the coordinate from the target to the center of
mass of the projectile, $L$ is the orbital angular momentum of the
projectile-target relative motion and $J$ is the total angular
momentum, $\vec{J}=\vec{L}+\vec{j}$.  The radial functions
$f_{Lnj}^J(R)$ satisfy the system of coupled equations
\begin{eqnarray}
& &\lc
 -\frac{\hbar^2}{2m_r}\lp\frac{d^2}{dR^2}-\frac{L(L+1)}{R^2}\rp+\varepsilon_{nj}-E\rc
 f_{Lnj}^J(R)\nonumber\\ &
 &+\sum_{L'n'j'}i^{L'-L}V^J_{Lnj,L'n'j'}(R)f_{L'n'j'}^J(R)=0,
\label{CCE3}
\end{eqnarray}
where $m_r$ is the reduced mass of the projectile-target system. The
coupling potentials $V^J_{Lnj,L'n'j'}(R)$ are then
\begin{equation}
V^J_{Lnj,L'n'j'}(R)=\la
LnjJM|\widehat{V}_{pt}(\vec{r}_1,\vec{r}_2,\vec{r}_3)|L'n'j'JM\ra,
\label{pff}
\end{equation}
where the ket $|LnjJM\ra$ denotes the function
$\Phi_{Lnj}^{JM}(\widehat{R},\vec{x},\vec{y})$ given by
\begin{equation}
\Phi_{Lnj}^{JM}(\widehat{R},\vec{x},\vec{y})=\sum_{\mu
M_L}\phi^{THO}_{nj\mu}(\vec{x},\vec{y})\la LM_Lj\mu|JM\ra
Y_{LM_L}(\widehat{R}).
\label{wfp}
\end{equation}

To calculate these coupling potentials, a multipole expansion of the
projectile-target interaction is developed. The procedure is analogous
to that for a three-body problem reported in Ref.~\cite{Nun99}. In
that work the traditional method of bin averaging was used as
discretization method instead of the THO method. We assume that the
projectile-target interaction is the sum of the interactions of each
particle of the projectile with the target, $V_{kt}(\vec{r}_k)$ with
$k=1,2,3$. For each pair potential, an appropriate Jacobi set is
chosen so that the corresponding coordinate $\vec{r}_k$ depends only
on the vectors $\vec{R}$ and $\vec{y}_k$. Assuming that the potentials
are central, the coefficients of the multipole expansion are generated
as
\begin{equation}
\mathcal{V}^k_{Q}(R,y_k)=\frac{1}{2}\int_{-1}^{+1} V^k(r_k)
P_Q(z_k)dz_k,
\label{potleg}
\end{equation}
where $P_Q(z_k)$ is a Legendre polynomial, $Q$ is the multipole order
and $z_k=\widehat{\vec{y}}_k\cdot\widehat{\vec{R}}$ is the cosine of
the angle between $\vec{y}_k$ and $\vec{R}$. So, the coupling
potential can be expressed as
\begin{widetext}
\begin{equation}
V^J_{Lnj,L'n'j'}(R)=\sum_{Q}(-1)^{J-j}\hat{L}\hat{L}'\left(\begin{array}{ccc}L&Q&L'\\0&0&0\end{array}\right)
W(LL'jj',QJ)F^Q_{nj,n'j'}(R),
\label{ef}
\end{equation}
\end{widetext}
where the radial form factor $F^Q_{nj,n'j'}(R)$ is
\begin{widetext}
\begin{eqnarray}
F^Q_{nj,n'j'}(R)&=&(-1)^{Q+2j-j'}\hat{j}\hat{j}'(2Q+1)\sum_{\beta\beta'}\sum_{k=1}^3\sum_{\beta_k\beta'_k}N_{\beta\beta_k}N_{\beta'\beta'_k}\nonumber\\
&\times&(-1)^{l_{xk}+S_{xk}+j'_{abk}-j_{abk}-I_k}\delta_{l_{xk}l'_{xk}}\delta_{S_{xk}S'_{xk}}
\hat{l}_{yk}\hat{l}'_{yk}\hat{l}_k\hat{l}'_k\hat{j}_{abk}\hat{j}'_{abk}
\left(\begin{array}{ccc}l_{yk}&Q&l'_{yk}\\0&0&0\end{array}\right)\nonumber\\
&\times&
W(l_kl'_kl_{yk}l'_{yk};Ql_{xk})W(j_{abk}j'_{abk}l_kl'_k;QS_{xk})W(jj'j_{abk}j'_{abk};QI_k)\nonumber\\
&\times&\int\int(\sin{\alpha_k})^2(\cos{\alpha_k})^2d\alpha_k~
\rho^5d\rho~ R_{n\beta
j}^{THO}(\rho)\varphi_{K_k}^{l_{xk}l_{yk}}(\alpha_k)\mathcal{V}^k_{Q}(R,y_k)\varphi_{K'_k}^{l_{xk}l'_{yk}}(\alpha_k)R_{n'\beta'j'}^{THO}(\rho),
\label{facfor}
\end{eqnarray}
\end{widetext}
with $\beta_k$ being the set of quantum numbers in the $k$'th Jacobi
system where the potential depends on $x_k$, and $\beta$ being the set
in the Jacobi system in which the states of the projectile are
calculated.  The matrix elements $N_{\beta\beta_k}$ transform the
hyperangular, angular and spin part of the wave functions from one
Jacobi set to another. Their explicit expression as a function of the
Raynal-Revai coefficients is developed in Ref.~\cite{face}. Note that
Eqs.~(\ref{ef}) and (\ref{facfor}) are completely general, and do not
depend on the nature of the basis.

\section{Structure model for $^6$He}

The $^6$He nucleus is treated here as a three-body system, comprising
an inert $\alpha$ core and two valence neutrons. The ground state has
total angular momentum $j^{\pi}=0^+$ with experimental binding energy
of $0.973$ MeV.  The ground state wave function was obtained by
solving the Schr\"odinger equation in hyperspherical coordinates,
following the procedure described in \cite{Zhu93,face}, and making use
of the codes
{\sc face} \cite{face}+ {\sc sturmxx} \cite{sturm}.  In these
calculations, the $n$-$^4$He potential was taken from
Ref.~\cite{Bang79}. It consists of an energy independent Woods-Saxon
potential, supplemented by a spin-orbit term with a Woods-Saxon
derivative radial shape.  This potential reproduces the low-energy s-
and p-phase shifts up to 10 MeV.  For the $NN$ interaction we used the
potential proposed by Gogny, Pires and Tourreil (GPT) \cite{gpt},
which contains central, spin-orbit and tensor components.  This
interaction was developed to give simultaneously an acceptable fit to
two nucleon scattering data up to 300 MeV and to describe reasonably
the properties for finite nuclei, particularly the radii, within the
Hartree-Fock approximation.  Besides the two-body ($n-n$ and
$n-\alpha$) potentials, the model Hamiltonian also includes a simple
phenomenological three-body force, depending only on the hyperradius,
according to the following power form
\begin{equation}
v_{3b}(\rho)=-\frac{a}{1+(\rho/b)^c},
\end{equation}
where $a$, $b$, and $c$ are adjustable parameters.  This potential is
introduced to correct the under-binding caused by our neglect of other
configurations, such as the $t$+$t$ channel.

We have performed different calculations that truncate the maximum
hypermomentum at $K_\mathrm{max}=2,4,6,8,$ and $10$, respectively.
For each value of $K_\mathrm{max}$, the three-body potential has been
adjusted to give the same binding energy and mean square radius 
(for $j^\pi=0^+$ states)
and the same position for the $2^+$ resonance (for $j^\pi=2^+$ states).
The latter value was also used for the $j^\pi=1^-$ states. 
The parameter $a$ varies with $K_\mathrm{max}$ and $j$, 
being of the order of 4 MeV for $j^\pi=0^+$ and 3 MeV for $j^\pi=1^-,2^+$.
The parameter $b$ varies with $K_\mathrm{max}$, within the range 4-6 fm. 
The parameter $c$ was fixed to 3 in all cases. 

The number of channels $\beta$ for each calculation increases
drastically with $K_\mathrm{max}$, making the calculations much more
demanding computationally.  In the following, unless stated otherwise,
the calculations presented use the basis with $K_\mathrm{max}=8$.  As
we will show below, this basis provides converged results with respect to
the hypermomentum for all the reactions considered in this work.
For this case, the number of channels $\beta$ is $15$ for
$j^{\pi}=0^+$, $26$ for $j^{\pi}=1^-$ and $46$ for $j^{\pi}=2^+$.  The
calculated three-body wave function has a binding energy of $0.95372$
MeV and a rms point nucleon matter radius of $2.46$ fm when assuming
an alpha-particle rms matter radius of $1.47$ fm.

\begin{figure}
\resizebox*{0.5\textwidth}{!}{\includegraphics{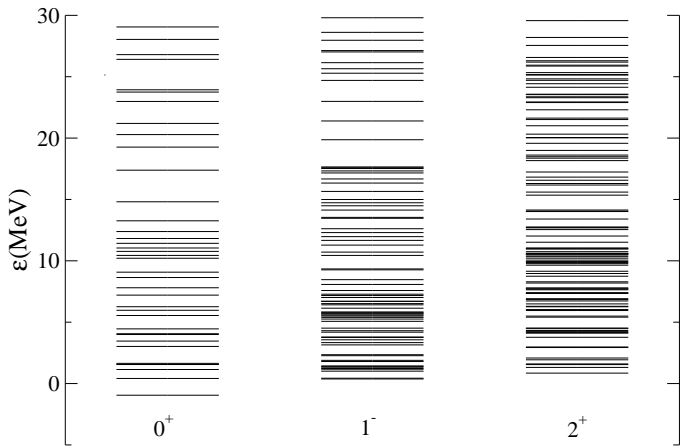}}
\vspace{-0.4cm} \caption{\label{ener} Energy spectrum of the states 
with  $j^{\pi}=0^+$, $j^{\pi}=1^-$, and $j^{\pi}=2^+$ up to 30 MeV excitation, 
obtained for a THO basis with $K_\mathrm{max}=8$ and $n_b=4$.} 
\end{figure}

The Jacobi set in which the two neutrons are related by the coordinate
$\vec{x}$ is chosen to generate the THO basis.  Applying the THO
method and diagonalizing the Hamiltonian in a finite THO basis, a set
of eigenstates is obtained. For $j=0^+$, the diagonalization produces
a state with negative energy, that corresponds to the ground state of
the system. The remaining eigenvalues appear at positive energy, and
are then associated with a discrete representation of the continuum
spectrum.
As an example, in Fig.~\ref{ener} we present the distribution of
eigenvalues obtained for a basis with $n_b=4$, for the states $j=0^+$,
$j=1^-$, and $j=2^+$ up to 30 MeV.

\section{Application to reactions}

In this Section we apply the formalism developed in Sec.~\ref{multexp}
to different reactions induced by $^6$He, taking the eigenstates
obtained with the THO basis to represent the projectile states. We
note that, even for a small value of $n_b$, the THO method will
produce eigenstates at very high excitation energies. States above a
certain excitation energy will not be relevant for the description of
the collision process, since they will be very weakly coupled. For
this reason, in these calculations the basis is truncated at a maximum
excitation energy, and only those eigenstates below this value were
included in the coupled-channels calculation. The maximum energy is
chosen independently for each reaction and each $n_b$, in order
to achieve convergence of the results with respect to this energy.

In the present calculations, only the $^6$He states with
$j=0^+,1^-,2^+$ are considered. Previous CDCC calculations for the
reactions studied in this work \cite{Mat04,Mat06,Mor07} have shown
that using these partial waves is sufficient to obtain converged
results and to describe satisfactorily the existing data for such
reactions.  We did not attempt to include higher partial waves, since
this would make the calculations very demanding computationally .

For these included $j^{\pi}$ states the coupling potentials given by
Eq.~(\ref{ef}) are calculated for multipolarities $Q=0,1,2$. Both
Coulomb and nuclear interactions were included. We emphasize that
continuum-continuum couplings were also included.  The diagonal as
well as non-diagonal coupling potentials were generated by folding the
neutron-target and $\alpha$-target interactions according to
Eq.~(\ref{pff}).  These interactions are represented by
phenomenological optical potentials at the relevant projectile
incident energy per nucleon \cite{Joh72}.  Then, the coupled equations
(\ref{CCE3}) are solved with the code {\sc fresco} \cite{Thom88}, that
reads the coupling potentials from external files.  
In most cases, the Numerov method was used to solve the coupled equations. 
However, in some cases, particularly when  excitation energies close
 to the total kinetic energy are involved, this method was found to be 
numerically unstable, and the R-matrix method~\cite{La58} was used instead. 
This 
method is more time consuming but has the advantage of being numerically 
more stable. 
In the following,
we present the results for different reactions for which experimental
data exist.

{\em{$^6$He$+^{12}$C}}. We study this reaction at 229.8~MeV, for
comparison with the experimental data of Lapoux et al.~\cite{Lap02}.
The $n+^{12}$C potential was taken from the global parametrization of
Watson et al.~\cite{Wat69}. The $\alpha+^{12}$C potential was
represented in terms of a standard Woods-Saxon shape with the
parameters adjusted in order to reproduce the elastic data for this
system at 34.75 MeV per nucleon \cite{Smi73}.  The parameters for
these potentials are listed in Table~\ref{tablepot}.

\begin{table}
\caption{\label{tablepot}Optical model parameters used in this work. All potentials are parametrized using the usual Woods-Saxon
form, with a real volume part and volume ($W_v$)  and surface ($W_d$)
imaginary part. Reduced
radii are related to physical radii by $R=r_0 A_T^{1/3}$.}
\begin{ruledtabular}
\begin{tabular}{c|ccccccc}
System &\( V_{0} \)&\( r_{0} \)&\( a_{0} \)&\( W_{v} \)&\( W_{d} \)&\( r_{i} \)&\( a_{i} \)\\
& (MeV) & (fm)  & (fm)  & (MeV)  & (MeV)  & (fm)  & (fm)  \\
\hline
$n$+$^{12}$C & 49.46 & 1.115 & 0.57 & 3.05 & 7.48 & 1.15 & 0.5 \\
$\alpha$+$^{12}$C & 100. & 1.289 & 0.71 & 19.98 & & 1.738 & 0.495\\
$n$+$^{64}$Zn & 51.82 & 1.203 & 0.668 & 0.29 &      & 1.203 & 0.668 \\
              &       &       &       &      & 5.97 & 1.279 & 0.534 \\
$\alpha$+$^{64}$Zn & 123 & 1.676 & 0.43 & 20.40 & & 1.467 & 0.43 \\
$n$+$^{208}$Pb & 47.37 & 1.222 & 0.726 & & 6.24 & 1.302 & 0.351 \\
$\alpha$+$^{208}$Pb & 96.44 & 1.376 & 0.625 & & 32. & 1.216 & 0.42 \\
\end{tabular}
\end{ruledtabular}
\end{table}

\begin{figure}\resizebox*{0.5\textwidth}{!}{\includegraphics{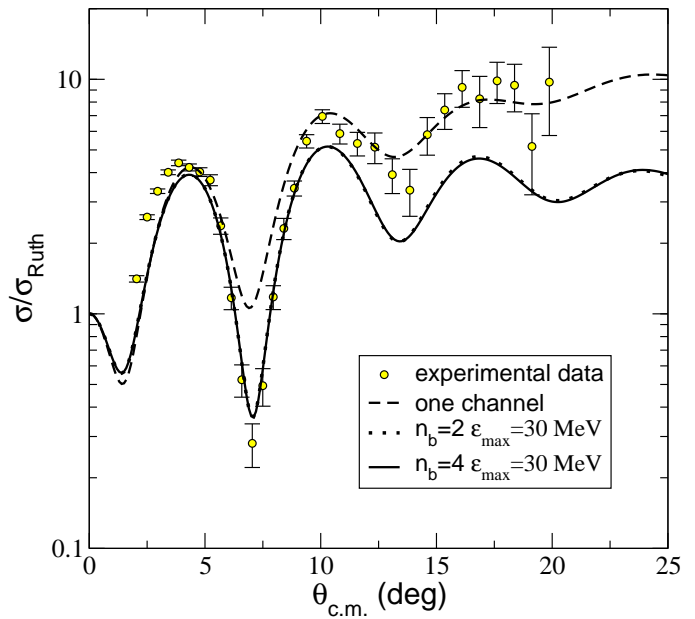}}
\vspace{-0.4cm} \caption{\label{he6c_el}(Color online)  Elastic differential cross section relative to Rutherford as a function of the scattering angle in the projectile-target center of mass for the reaction $^6$He$+^{12}$C at $E_\mathrm{lab}$=229.8 MeV. 
Experimental data are from Ref.~\cite{Lap02}.}
\end{figure}

The coupled equations were solved up to $J=70$ and the solutions were
matched to their asymptotic form at the radius $R_m=200$ fm.  In
Fig.~\ref{he6c_el} we present the angular distribution of the elastic
differential cross section relative to Rutherford. The thick solid
line is the full CDCC result for a basis with $n_b$=4.  This
calculation reproduces the data fairly well (open circles) up to
10$^{\circ}$, but it clearly underestimates the data points at larger
angles.  Interestingly, this effect was also found in the
phenomenological analysis of Lapoux et al.~\cite{Lap02}, as well as in
the four-body CDCC calculation of Matsumoto et al.~\cite{Mat04} for
the same reaction.  We also show the analogous calculation when
omitting all the couplings to the continuum (one channel calculation)
with a dashed line. For the reaction at $229.8$ MeV we conclude that
the effect of coupling to the continuum is a reduction of the cross
section for angles beyond 5$^{\circ}$. This effect has also been
observed in the scattering of $^{11}$Be+$^{12}$C at $E\simeq 49$~MeV
per nucleon \cite{Ron97b}, and is probably present in other reactions
induced by weakly bound projectiles at energies of a few tens of MeV
per nucleon.  That the no-continuum coupling calculation reproduces
the data reasonably well at the larger angles is probably fortuitous,
and cannot be attributed to the adequacy of this approximation.  As we
have shown, continuum couplings are very important in this reaction.

We also show in Fig.~\ref{he6c_el} the full CDCC calculation for
$n_b$=2 (dotted line).  This calculation is practically
undistinguishable from the calculation with $n_b$=4, indicating that
it is not necessary in this case to have a very precise discretization
of the continuum in terms of excitation energy.
We found that a maximum excitation energy of
$\varepsilon_\mathrm{max}=$30~MeV provided good convergence for all
the values of $n_b$ presented.

{\em{$^6$He$+^{64}$Zn}}. We have studied this reaction at two
different energies, namely, 13.6~MeV and 10~MeV, for which
experimental data exist \cite{Dip04}. The $n+^{64}$Zn potential was
taken from the global parametrization of Koning and Delaroche
\cite{Kon03}. For the $\alpha+^{64}$Zn system, we took the optical
potential derived in Ref.~\cite{Dip04}. The parameters are listed in
Table~\ref{tablepot}.  The coupled equations were solved up to $J=60$
and 40, respectively, and for projectile-target separations up to
$R_m=100$~fm.

\begin{figure}\resizebox*{0.5\textwidth}{!}{\includegraphics{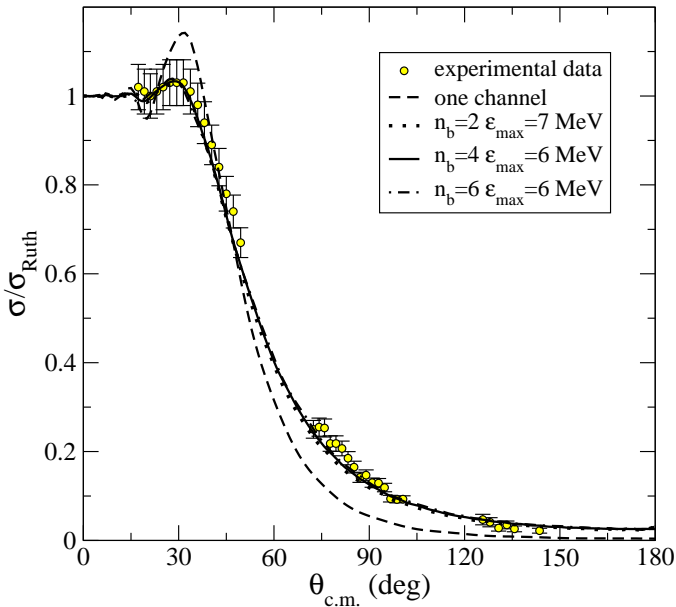}}
\vspace{-0.4cm} \caption{\label{he6zn_e13.6_el}(Color online) Elastic differential cross section relative to Rutherford as a function of the c.m.\ scattering angle for the reaction $^6$He$+^{64}$Zn at $E_\mathrm{lab}$=13.6 MeV. Experimental data are from Ref.~\cite{Dip04}.}
\end{figure}

\begin{figure}\resizebox*{0.5\textwidth}{!}{\includegraphics{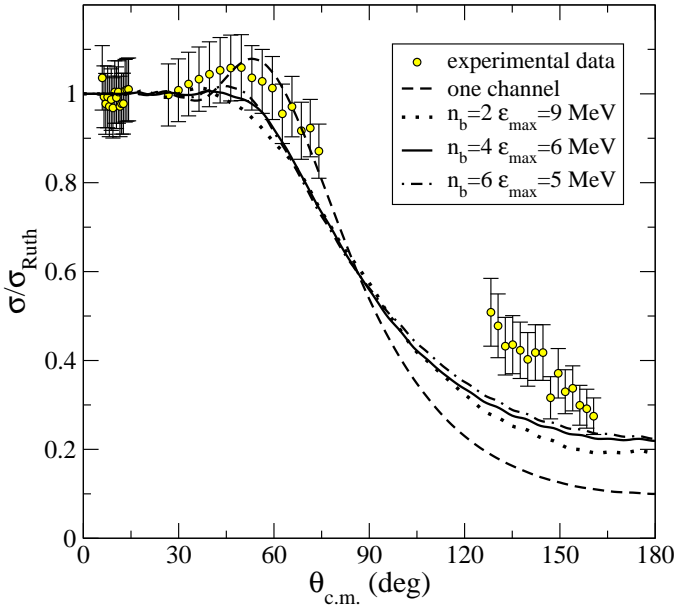}}
\vspace{-0.4cm} \caption{\label{he6zn_e10_el}(Color online) Elastic differential cross section relative to Rutherford as a function of the c.m.\ scattering angle for the reaction $^6$He$+^{64}$Zn at $E_\mathrm{lab}$=10 MeV. Experimental data are from Ref.~\cite{Dip04}.}
\end{figure}

In Figs.~\ref{he6zn_e13.6_el} and \ref{he6zn_e10_el} we present the
experimental and calculated angular distributions of the elastic cross
section for these two reactions.  The dashed lines correspond to the
one channel calculations (i.e., omitting the continuum) 
and the thick solid lines are the full four-body CDCC calculations for 
a basis with $n_b=$4.

At $E_{\mathrm{lab}}$=13.6 MeV (Fig.~\ref{he6zn_e13.6_el}), 
the one-channel calculation exhibits a
pronounced rainbow peak at around 30$^{\circ}$, which is much smaller
in the data. Also, this calculation gives a too small cross section at
large angles.  Inclusion of couplings to the continuum suppresses this
rainbow, and enhances the backward angles cross section, improving the
agreement with the data in the whole angular range. In the same
figure, we also show the full CDCC calculation for 
a basis with $n_b$=2 (dotted) and 6 (dot-dashed). These two calculations are
very close to $n_b$=4 showing a very good convergence with respect
$n_b$.
The maximum excitation energy required for convergence depended
somewhat on the value of nb, ranging from $\varepsilon_\mathrm{max}$=7
MeV (for nb=2) to $\varepsilon_\mathrm{max}$=6 MeV (for nb=6)

At $E_{\mathrm{lab}}$=10 MeV (Fig.~\ref{he6zn_e10_el}), 
the full CDCC calculation also
improves the agreement with the data at backward angles, although some
underestimation remains.  Interestingly, the data suggests the
presence of a rainbow at around 50$^{\circ}$, which is not present in
our calculation. It should be noted that the experimental error bars
are large at this energy, so more accurate measurements would be
needed to make more definite conclusions about this apparent
discrepancy. Again, in the same figure, we show
the full CDCC calculation for a basis with $n_b$=2 (dotted) and 6
(dot-dashed).  In this case, we find that the convergence with
respect $n_b$ is slower.  However the calculations with $n_b$=4 and 6
are quite close and give a reasonable convergence.
As in the previous case, 
the maximum excitation energy required for convergence depended
somewhat on the value of nb, ranging from $\varepsilon_\mathrm{max}$=9
MeV (for nb=2) to $\varepsilon_\mathrm{max}$=5 MeV (for nb=6).

{\em{$^6$He$+^{208}$Pb}}. We have performed calculations for this
reaction at 22~MeV, in order to compare with the recent data of
S\'anchez-Ben\'itez et al.~\cite{Benln}. We took the $n+^{208}$Pb
potential from Ref.~\cite{Rob91} and the $\alpha+^{208}$Pb potential
from Ref.~\cite{Bar74}. The parameters for these potentials are also
listed in Table~\ref{tablepot}.  The coupled equations were solved up
to $J=150$ and matched to their asymptotic solution at $R_m=200$ fm.

\begin{figure}\resizebox*{0.5\textwidth}{!}{\includegraphics{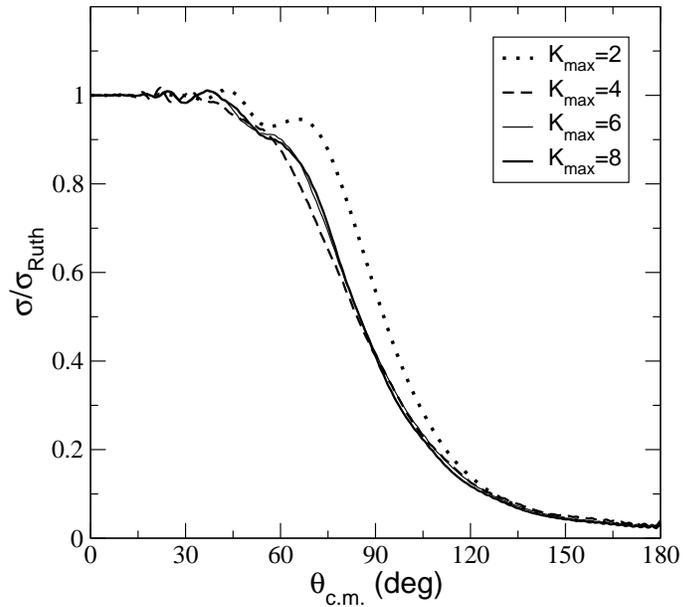}}
\vspace{-0.4cm} \caption{\label{he6pb_el_conK} Convergence of 
the differential elastic cross section
with respect to $K_\mathrm{max}$, for the reaction $^6$He$+^{208}$Pb at $E_\mathrm{lab}$=22 MeV. 
All the calculations use $n_b=4$ for the number of hyperradial excitations 
and the maximum excitation energy was set to 8 MeV.}
\end{figure}

First, we discuss the convergence of the calculation with respect
the hypermomentum ($K_\mathrm{max}$) and the hyperradial excitation
($n_b$).  In Fig.~\ref{he6pb_el_conK}, we show the calculations
with different values of $K_\mathrm{max}=2,4,6,8$ and for the same
value of $n_b$=4. For a meaningful comparison, in all these cases the
three-body potential was adjusted in order to give the same binding energy and
rms radius, for $j=0^+$ and the same position for the resonance, for
$j=2^+$. We found a relatively fast convergence with respect to this
parameter. In particular, the calculations with $K_\mathrm{max}=6$,8
and 10 are very similar (for clarity, the latter has been omitted from
the figure).
For rest of reactions the results are quite similar, achieving the
convergence for $K_\mathrm{max}=$6 or 8.

\begin{figure}\resizebox*{0.5\textwidth}{!}{\includegraphics{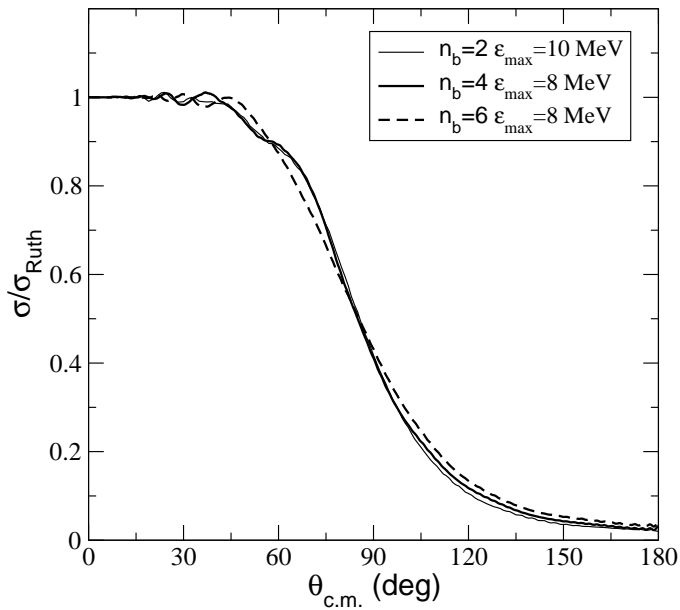}}
\vspace{-0.4cm} \caption{\label{he6pb_el_connb}  Convergence of 
the differential elastic cross section
with respect to $n_b$, for the reaction $^6$He$+^{208}$Pb at $E_\mathrm{lab}$=22 MeV.}
\end{figure}

The convergence with respect to $n_b$ for this reaction is illustrated in
Fig.~\ref{he6pb_el_connb}. For clarity, we show only the results for
even values of $n_b$.  Unlike the previous cases, the convergence rate
found in this case was rather slow. Although the differences in the
calculated cross sections are less than 5\%, the oscillatory pattern
at the rainbow region changes from one value of $n_b$ to another.  A
possible explanation for this slow convergence rate is given below.

\begin{figure}\resizebox*{0.5\textwidth}{!}{\includegraphics{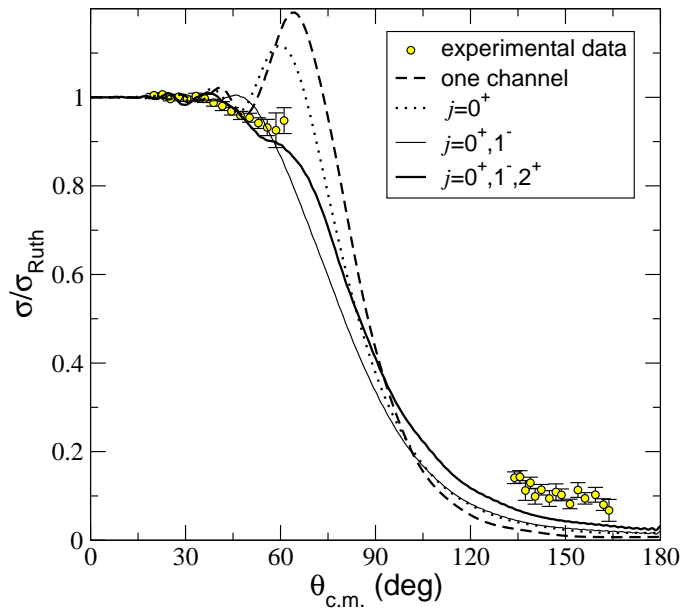}}
\vspace{-0.4cm} \caption{\label{he6pb_el} (Color online) Elastic differential cross section relative to Rutherford as a function of the scattering angle in the projectile-target center of mass for the reaction $^6$He$+^{208}$Pb at $E_\mathrm{lab}$=22 MeV. The full CDCC calculations uses $n_b=$4 and $\varepsilon_\mathrm{max}=8$ MeV. Experimental data are from Ref.~\cite{Benln}.}
\end{figure}

In Fig.~\ref{he6pb_el} we compare the experimental and calculated
angular distributions of the elastic cross section.  The dashed line
is the one channel calculation and the thick solid line the full
CDCC calculation including the continuum.  The latter uses
$K_\mathrm{max}=8$, $\varepsilon_\mathrm{max}=8$ MeV, $n_b=4$.  The
one channel calculation shows a rainbow that disappears in the full
calculation, in agreement with the data. At backward angles, the
agreement with the data is improved when we include the coupling to
the continuum. In order to show the contribution of the couplings to each $j$,
we also include in this figure the calculation including only $j=0^+$ 
states (dotted line) and the calculation  with $j=0^+,1^-$ states (thin
 solid line). From
these calculations we can conclude that dipole couplings are the main
responsible for the characteristic reduction of the cross section at
the angles around the rainbow. The strong influence of dipole
couplings might explain the slow convergence with respect to the
parameter $n_b$ found for this reaction. These couplings are very
sensitive to the excitation energy of dipole states, which appear at
different positions in our discrete representation of the $^6$He
continuum, as we vary the number of hyperradial excitations, $n_b$. 
By
contrast, in the $^6$He+$^{12}$C case, dipole excitations are very
small, and this might explain the fast convergence with respect to
$n_b$ in that case.

Moreover, 
we find that the range of the form factors [Eq.~(\ref{facfor})]
changes significantly for the different pseudo-states as $n_b$ is changed. 
This could also contribute to the slow convergence at scattering energies 
close to the Coulomb barrier.


\section{Summary and conclusions}

The collision of a loosely bound three-body projectile with a target
nucleus has been studied in the framework of the continuum-discretized
coupled-channels (CDCC) method. A set of normalizable states, also
known as pseudo-states, is used to represent the three-body continuum
of the projectile. In particular we took the Transformed Harmonic
Oscillator (THO) basis, which is constructed from the ground state of
the system.  Within the spirit of the CDCC approach, a multipole
expansion of the coupling potentials has been developed for a
four-body system (three-body projectile plus a target).
 
The formalism has been applied to the reactions $^6$He$+^{12}$C at
$229.8$ MeV, $^6$He$+^{64}$Zn at $13.6$ and $10$ MeV, and
$^6$He$+^{208}$Pb at $22$ MeV, taking into account both the Coulomb
and nuclear interactions.

Overall, we find good agreement between the calculated and the
experimental elastic scattering angular distributions.  However, for
the $^6$He$+^{12}$C reaction at $229.8$ MeV the calculations
underestimate the experimental data for c.m.\ scattering angles beyond
10$^\circ$.  The fact that this effect was
also found in previous analyses of this reaction \cite{Lap02,Mat04}
suggests that the discrepancy is not related to the particular
features of our approach.

For the reaction $^6$He$+^{64}$Zn at $13.6$ and $10$ MeV the
calculations are in fair agreement with the data, the reproduction
being better in the higher energy case. At $10$ MeV our calculations
do not predict a rainbow at around $50^\circ$, a hint of which is seen
in the data, but is broadly consistent with the data within the stated
experimental errors.

In actual coupled-channels calculations, the discrete basis has to
be truncated in the excitation energy ($\emax$), the maximum
hypermomentum ($K_\mathrm{max}$), and the maximum number of
hyperradial excitations ($n_b$). In all the cases under study, we have
found a good convergence of the calculated observables with respect to
the parameters $\emax$ and $K_\mathrm{max}$. However, the rate of
convergence with respect to $n_b$ was found to depend very much on the
specific reaction. For the reaction $^6$He$+^{12}$C at $229.8$ MeV the
convergence was found to be very fast, with $n_b=2$ providing fully
converged results. For $^6$He$+^{64}$Zn at near-barrier energies, we
required $n_b \approx 4$ for an acceptable convergence. Finally, for
$^6$He$+^{208}$Pb at 22 MeV, the convergence was found to be slow and
oscillatory. In fact, our biggest calculation, corresponding to
$n_b=6$, is still not fully converged.  
Because of computational limitations we have not explored
this question further,
as required to study the convergence
of the calculations with respect to the basis size.



This work shows that the use of the transformed harmonic oscillator
basis, developed in previous works, combined with the standard CDCC
method, provides a reliable procedure for the treatment of the
scattering of a loosely bound three-body projectile by a target.  It
will be interesting to compare this method with other representations
of the continuum, including the standard discretization procedure in
terms of continuum bins which, in the case of three-body projectiles,
requires the calculation of the three-body scattering states. This
work is underway and the results will be published elsewhere.

\begin{acknowledgments}
This work was supported in part by the DGICYT under
projects FIS2005-01105 and FPA 2006-13807-C02-01, and
in part by the FCT under the grants POCTI/ISFL/2/275 and
POCTI/FIS/43421/2001.  
This work was performed under the auspices of the U.S. Department 
of Energy by Lawrence Livermore National Laboratory under Contract 
DE-AC52-07NA27344.
J.A.T. acknowledges the support of the the United Kingdom Science and Technology Facilities Council (STFC) under Grant No.~EP/D003628.
A.M.M. acknowledges a research
grant from the Junta de Andaluc\'ia. M.R.G. acknowledges a
research grant from the Ministerio de Educaci\'on and support from the Marie
Curie Training Site program.
\end{acknowledgments}


\bibliographystyle{apsrev}
\bibliography{./4bCDCC}

\end{document}